\documentstyle[11pt,newpasp,twoside]{article}
\def\grtsim{{_ >\atop{^\sim}}}

\def\edcomment#1{\iffalse\marginpar{\raggedright\sl#1\/}\else\relax\fi}
\reversemarginpar

\begin{document}
\title{Formation and evolution of disk galaxies within cold dark matter halos }
\author{Vladimir Avila-Reese, Claudio Firmani\altaffilmark{1}}
\affil{Instituto de Astronom\'\i a-UNAM, A.P. 70-264, 04510 
M\'exico, D. F.}
\author{Anatoly Klypin and Andrey V. Kravtsov\altaffilmark{2}}
\affil{Astronomy Department, NMSU, P.O. Box 30001/Dept. 4500, Las Cruces, 
NM 88003-8001, USA}

\altaffiltext{1}{Also Osservatorio Astronomico di Brera, via E.Bianchi 
46, I-23807 Merate, Italy}

\altaffiltext{2}{Hubble Fellow. Current address: Department of Astronomy, 
The Ohio State University, 140 West 18th Av., Columbus, OH 43210-1173, USA}

\begin{abstract}
  We  present results of  extensive model calculations  of disk galaxy
  evolution within an  hierarchical inside-out formation scenario.  We
  first   compare   properties of  the  dark   halos   identified in a
  cosmological N-body simulation  with predictions of a  seminumerical
  method   based on  an extended    collapse model  and  find a   good
  agreement.  We then describe detailed modelling of the formation and
  evolution of disks within   these growing halos and  predictions for
  the   main properties,   correlations and evolutionary   features of
  normal  disk   galaxies.  The  shortcomings   of   the scenario  are
  discussed.
 
\end{abstract}

\section{Introduction}

The hierarchical   cosmic structure  formation picture based    on the
inflationary  cold  dark matter  (CDM) provides  a solid framework for
models  of galaxy  formation and evolution.   On  the other  hand, the
unprecedented observations of  galaxies at different redshifts make it
possible to probe and constrain these models.  Here we discuss some of
the  results obtained with  a self-consistent scenario  of disk galaxy
formation and evolution within the context of the hierarchical picture
(the extended collapse scenario).

\section{Dark matter halos}

Using the extended Press-Schechter approximation, we generate the mass
aggregation histories (MAHs) of  the dark matter (DM) halos.  Collapse
and   virialization  of  these   halos  are  then calculated  assuming
spherical symmetry and adiabatic invariance, using a method based on a
generalization of the secondary infall model (Avila-Reese, Firmani, \&
Hern\'andez  1998).  These halos  will  mainly  correspond to isolated
systems. The diversity  of MAHs results in  {\it diversity} of density
profiles which, in  our model, mainly depend on  the MAH.  The density
profile corresponding  to the  average MAH  is  well described by  the
Navarro et al. (1996)  profile. In Avila-Reese  et al. (1999)  we have
compared  the outer density   profiles, concentrations and  structural
relations  of  thousands  of  halos    identified  as  isolated in   a
cosmological ($\Lambda$CDM) N-body simulation with those obtained with
our seminumerical method. We found a  good agreement between the model
and simulation results.

We   have found  that $\sim  13\%$    of the halos    in the numerical
simulation at $z=0$ are contained within larger  halos and $\sim 17\%$
have significant companions  within three virial radii.  The remaining
70\% of the halos are isolated objects. The slope $\beta$ of the outer
density profile ($\rho \propto r^{-\beta}$) and the halo concentration
defined as  $c_{1/5}=r_h/r(M_h/5)$,  where  $r_h$ and $M_h$    are the
virial  radius and mass,  depend on the  halo environment. For a given
$M_h$, halos in clusters have typically steeper outer profiles and are
more  concentrated   than the isolated   halos  (for the latter $\beta
\approx 2.9$  in average and $\beta$  between 2.5 and  3.8 for 68\% of
the halos). Contrary to naive expectations,  halos in galaxy and group
systems  as   well    as  the  halos   with    significant companions,
systematically have flatter and   less concentrated density   profiles
than isolated halos. A tight correlation between $M_h$ and the maximum
circular velocity $V_{m}$  is observed: $M_h\propto  V_m^n$, $n\approx
3.2$. This is roughly the slope of the infrared Tully-Fisher relations
(TFR).  Thus, it seems  that there is no  room for the mass dependence
of the infrared $M_h/L$ ratio.

\section{Galaxy evolutionary models}

We  model the formation and  evolution of baryon  disks in centrifugal
equilibrium within the growing CDM halos formed as described  in \S 2. 
We assume that halos acquire angular momentum from large-scale torques
with the   spin   parameter $\lambda$  distributed   log-normally  and
constant in time.  The disks are built  inside-out with the gas infall
rate (no mergers)  proportional  to the cosmological mass  aggregation
rate and    assuming detailed   angular  momentum  conservation.   The
gravitational drag of the disk on the DM halo is calculated. The local
SF is  assumed to  be induced by  disk instabilities  and regulated by
energy balance  within the disk   turbulent  ISM (no SF  feedback  and
self-regulation at the level of  the interhalo medium is allowed).  We
also  calculate the secular formation of  a bulge.  This  way, {\it at
  each epoch and at each radius}, the growing disk is characterized by
the infall rate of  fresh  gas, the gas  and stellar  surface  density
profiles, the total rotation  curve (including the DM  component), the
local SF rate, and the size of the inner region transformed into bulge
component.

\section{Highlights of the model results}

Results on the structure and dynamics of  our model disk galaxies were
discussed in Firmani \&  Avila-Reese (2000); the luminosity properties
and  topics   related to the disk    Hubble sequence  were  treated in
Avila-Reese \& Firmani (2000), while  some evolutionary aspects of the
galaxies  were   presented in  Firmani \& Avila-Reese   (1999). In the
following, we highlight some of the results.

{\bf Local properties.}  The (stellar) surface density and  brightness
profiles are  exponential,  the   sequence of  high to   low   surface
brightness (SB) being mainly determined by $\lambda$. The gas profiles
at $z=0$ are also exponential although much lower  in density and with
a scale radius  $\sim 2-4$  times larger  than  the stellar profiles.  
There is a negative  radial gradient of the color  index: stars in the
outer regions of the disk form later than  stars in the inner regions. 
We find that the local SF  rate per unit area  correlates with the gas
surface density  as $\Sigma  _{\rm SFR}(r)\propto \Sigma_g^n(r)$  with
$n\approx 2$  for most of  the models and over a  major portion of the
disks.   The  shape  of the  rotation  curves  correlates with the  SB
($\lambda$) and  in most  cases is  approximately flat. The  dark halo
dominates in the rotation  curve  decomposition down to  very  central
regions.

{\bf The  infrared Tully-Fisher  relations  (TFR).}  The slope of  the
$M_h-V_m$ relation of  the CDM halos remains imprinted  in the TFR and
agrees  with observations. This   slope is almost  independent of  the
assumed  disk   mass fraction  $f_d$ when   the disk  component in the
rotation curve decomposition is gravitationally important ($f_d\grtsim
0.03$ for  the $\Lambda$CDM model used here).   The  zero point of the
model TFR is only  slightly larger than the observed  zero point.  The
rms scatter in our TFR slightly decreases with  mass; from $V_m=70$ to
300 km/s the scatter is between 0.38 and 0.31 mag.  We have found that
a major contribution  to this scatter  is from the  scatter in the  DM
halo structures   due  to  the dispersion  of     the MAHs;  a   minor
contribution to the scatter is due to the dispersion of $\lambda$. The
TFR for high and low SB models is approximately the same. The slope of
the  correlation among the residuals of  the  TF and luminosity-radius
relations  is small  and   non-monotonic, although  the shape  of  the
rotation  curves of  our  models correlates with the  SB.  For a given
total  (star+gas) disk mass,  the $V_m$ decreases  with decreasing SB. 
However, owing  to  the dependence of  the  SF efficiency on the  disk
surface density, the stellar  mass $M_s$ (luminosity) also  decreases. 
This combined influence of the SB  ($\lambda$) on $V_m$ and $M_s$ puts
models  of different SB on the  same  $M_s-V_m$ relation. As a result,
high and low SB models follow similar TFRs.

{\bf The Hubble sequence. } The main properties of the high and low SB
disk galaxies and their correlations are determined by the combination
of three fundamental physical factors  and their dispersions: the halo
virial mass, the MAH and the angular momentum given through $\lambda$.
The   MAH determines mainly  the  halo  structure, the integral  color
index, and the gas   fraction $f_g$ while $\lambda$  determines mainly
the disk   SB, the  bulge-to-disk (b/d)  ratio  and the  shape of  the
rotation curve. Our models show that  the redder and more concentrated
(higher SB)  is the disk,  the smaller is  $f_g$ and the larger is the
b/d ratio (disk  Hubble sequence). The values  of all these magnitudes
are in good agreement with observations.

{\bf  Evolutionary  features.}   In  the inside-out  hierarchical disk
formation  scenario galaxies   undergo  not only luminosity  but  also
structural (size, SB, b/d ratio) evolution. For an Einstein- de Sitter
universe we find that  the scale  radius for normal disk galaxies
decreases roughly as  $(1+z)^{-0.5}$ up to  $z\approx 1.5$, while  the
central $B-$band SB from $z=0$ to $z=1$ increases by $\approx 1.2$ mag.

The SF  history in the models  is driven both  by the MAH and the disk
gas  surface density. For the  average MAH  and $\lambda=0.05$, the SF
rate reaches a  maximum  at $z\approx 1.5-2.5$   which is a factor  of
2.5-4.0 higher   than the  rate at $z=0$.    In the same    way, $L_B$
increases towards the past   by factors slightly  smaller than  the SF
rate.  The  less massive  galaxies  present  a slightly  more   active
luminosity evolution than the massive galaxies. The model galaxies are
somewhat bluer in the past; from $z=0$ to $z=1$ the $B-V$ decreases on
average  0.25-0.30   magnitudes.  The total   mass-to-$L_B$ ratio also
decreases towards  higher redshifts: from $z=0$  to $z=1$ it decreases
on average  by a factor $\sim  3.3$,  i.e. a galaxy   at $z=1$ is more
luminous in the  $B-$band and less  massive than at $z=0$. Again, this
is a result related to the hierarchical MAHs of the protogalaxies.

Owing to  the mass  (size) evolution, {\it   for a fixed   $V_m$}, the
$H-$band luminosity  is a factor $\approx $2.2  less  at $z=1$ than at
$z=0$;  however, owing to the luminosity  evolution, $L_B$ is a factor
$\approx$ 2.1 larger. Therefore,  while the zero-point of the $H-$band
TFR increases towards  the  past, in the   case of  the  $B-$band TFR,
compensation due to the    $L_B$ evolution results in   the zero-point
remaining  approximately constant with  time. The slopes in both cases
also remain constant.

\section{Potential difficulties of the hierarchical scenario}

Although several main properties, correlations, and evolutionary
features of normal disk galaxies have been successfully predicted 
by our models, it is important to remark on their problems.
We find the following potential conflicts with the observations:
{\bf 1)} the size and SB evolution of the disks is too pronounced, {\bf 2)} 
the radial color index gradients are too steep and the $f_g$ is 
slightly over-abundant, {\bf 3)} the DM component dominates in the 
rotation curve decompositions almost down to the center and the 
halos are too cuspy. 

Regarding item 1),  if selection effects in the  deep field are not so
significant as Simard et al. (1999) have  claimed, then probably it is
not so serious.  In fact,  some physical ingredients not considered in
our   models   (e.g.,    merging,  angular   momentum  transfer,   and
non-stationary  SF)  all  work  in  the  direction  to  improve models
regarding problems 1) and 2). The problem 3) can probably be solved if
the  inner density  profile of  the  CDM halos can   be shallower than
predicted  (several solutions such as   self-interacting CDM, warm DM,
non-Gaussian fluctuations,   have been proposed). Nevertheless,  it is
possible that all these problems together  with those of the dearth of
satellites and the  high frequency of disk  disruptive mergers, are in
general  pointing out to serious  problems for  the Gaussian CDM-based
hierarchical picture of structure formation.  More observational tests
regarding the problems mentioned above  and more theoretical effort in
modeling galaxy formation and evolution are urgently required.

\end{document}